# Simultaneous Super-Resolution of Spatial and Spectral Imaging with a Camera Array and Notch Filters

Peng Lin, Xuesong Wang, Yating Chen, Xianyu Wu, Feng Huang, Shouqian Chen

**Abstract**—This study proposes an algorithm based on a notch filter camera array system for simultaneous super-resolution imaging and spectral reconstruction, enhancing the spatial resolution and multispectral imaging capabilities of targets. In this study, multi-aperture super-resolution algorithms, pan-sharpening techniques, and spectral reconstruction algorithms were investigated and integrated. The sub-pixel level offset information and spectral disparities among the 9 low-resolution images captured by the 9 distinct imaging apertures were utilized, leading to the successful reconstruction of 31 super-resolution spectral images. By conducting simulations with a publicly available dataset and performing qualitative and quantitative comparisons with snapshot coded aperture spectral imaging systems, the experimental results demonstrate that our system and algorithm attained a peak signal-to-noise ratio of 35.6dB, representing a 5dB enhancement over the most advanced snapshot coded aperture spectral imaging systems, while also reducing processing time. This research offers an effective solution for achieving high temporal, spectral, and spatial resolution through the utilization of multi-aperture imaging systems.

**Index Terms**—Camera array, super-resolution, computational imaging, spectral reconstruction, notch filter.

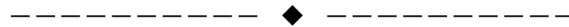

## 1 INTRODUCTION

Rapid acquisition of high-resolution three-dimensional spectral data cubes (spectral information, spatial information, and intensity information) from dynamic scenes holds significant application value in fields such as medical diagnosis [1], environmental monitoring [2], food inspection [3], and remote sensing classification [4]. Due to two-dimensional sensors can not directly capture three-dimensional spectral data cubes, existing spectral imaging technologies often necessitate a compromise between temporal resolution, spatial resolution, and spectral resolution. The most typical scanning spectral imaging systems [5],[6],[7] require scanning along the spatial or spectral dimensions, rendering them unfit for dynamic scenes. To solve the problem of rapid acquisition of high-resolution three-dimensional spectral data blocks in dynamic scenes, more and more research has focused on snapshot computational spectral imaging systems in recent years. Multi-spectral cameras based on micro-filter arrays integrate filter units on the detector surface and improve spectral resolution by sacrificing spatial resolution. However, most existing demosaicking algorithms are not applicable to a large number of spectral bands [8]. To elevate the spatial resolution of three-dimensional spectral data blocks, one typical approach involves incorporating a high-resolution panchromatic image or an RGB image for resolution enhancement. This image is then fused with the low-resolution spectral image to reconstruct a high-resolution multispectral image [9],[10],[11],[12]. However, the additional collection of a high-resolution panchromatic or RGB image implies an increase in the volume and weight of the optical system. The coded aperture snapshot spectral imaging systems(CASSI) [13] realize the encoding of spectral information of target scenes through dispersive elements and coded apertures. However, the coded aperture destroys the spatial information of the scene, leading to low spatial resolution of the reconstructed image and blurred details. Wang et al. [14] added a panchromatic detector to CASSI to obtain a panchromatic image rich in detail information, and their algorithm achieved good imaging results in the reconstructed spectral image but this method augments the system's structural complexity. Microlens arrays [15],[16] and camera array imaging systems [17] achieve spectral imaging by covering different filters in front of each aperture and combining computational imaging algorithms. However, the number of spectral bands that these imaging systems can capture is restricted by the number of apertures. Chen et al. [18] innovatively combined the multi-aperture super-resolution algorithm with a bandpass filter camera array system. They performed super-resolution reconstruction on the low-resolution bandpass spectral images captured by each aperture, achieving spatial super-resolution imaging of spectral images for the first time. By substituting bandpass filters with notch filters, Wang et al. [19] overcame the low light energy

- *P Lin. is with the School of Mechanical Engineering and Automation, Fuzhou University, Fuzhou 35018, China, and the School of Astronautics, Harbin Institute of Technology. Heilongjiang, Harbin 150001, China. E-mails: linpengfzu@sina.com*
- *R Cao. B Zhou. X Wang. Y Chen. X Wu, F Huang are with the School of Mechanical Engineering and Automation, Fuzhou University, Fuzhou 35018, China.*
- *E-mails: cao_rj@189.cn, bzhou0060@163.com, wxs2019gzyx@163.com, 13275911281@163.com, xwu@fzu.edu.cn, huangf@fzu.edu.cn.*
- *S Chen is with the School of Astronautics, Harbin Institute of Technology. Heilongjiang, Harbin 150001, China. E-mail:*
- *Corresponding author: Xianyu Wu, Feng Huang, Shouqian Chen.*

efficiency problem inherent to the spectral imaging system based on bandpass filters. They proposed a spectral super-resolution algorithm, realizing high spectral resolution imaging using notch filters. However, there remains significant room for enhancing the accuracy of the reconstructed spectral images by this algorithm. Wu et al. [20],[21] proposed a high spatial and temporal resolution imaging method based on notch filters and camera arrays. This method not only achieves high-quality imaging results but also enables higher spectral resolution image reconstruction.

The study delineated in reference [20] proposed a spectral reconstruction algorithm for multi-aperture multispectral imaging systems, aimed at improving the spectral resolution of the collected image data. However, the spatial resolution of the image was not improved in the process of computational reconstruction of the acquired spectral images. Intriguingly, images collected by different apertures in camera array imaging systems inherently exhibit sub-pixel offsets, a feature that has been proven to be advantageous for the computational reconstruction of images with superior spatial resolution [22]. As a result, the super-resolution imaging algorithm for multi-frame, multi-aperture images has the potential to be applied to multi aperture multispectral imaging systems, facilitating the computational reconstruction of spectral images with high spatial resolution. An imaging system using a camera array equipped with bandpass filters collects inconsistent image bands, engendering significant spatial structural variations and brightness discrepancies between images from different apertures, making them unsuitable for multi-aperture super-resolution computational imaging. In [18], a bandpass system with nine apertures was employed, generating spectral images for eight other bands per aperture through a spectral clustering methodology. Despite the high spectral accuracy of these generated spectral images, they introduce spatial errors, thereby impacting the precision of super-resolution reconstruction. Contrastingly, images captured using notch filters closely resemble panchromatic images, and their application in super-resolution image reconstruction serves to address this challenge. Hence, this paper introduces an innovative algorithm, Spatial and Spectral Super-Resolution Reconstruction (SSR), designed for the notch-filtered and panchromatic images acquired by the proposed camera array imaging system. By harnessing both the notch-filtered and panchromatic images, SSR concurrently improves the spatial and spectral resolutions of the acquired images. Primarily, a near-panchromatic high-resolution image is procured via the multi-aperture super-resolution algorithm employing maximum likelihood estimation [22]. Subsequently, the resolution of the notch image for each aperture is enhanced through a panchromatic sharpening algorithm [23]. Ultimately, high-resolution spectral information of the target scene is reconstructed via the spectral reconstruction algorithm proposed in this paper, thereby achieving the computational reconstruction of high-resolution multispectral images from nine-aperture low-resolution notch spectral images. Through both simulation and experimental verification, this paper ascertains that our proposed methodology and algorithm effectively bolster spatial super-resolution and spectral fidelity, without engendering image artifacts.

## 2 SPATIAL AND SPECTRAL SR IMAGING PRINCIPLES OF THE NOTCH FILTERS BASED MULTI-APERTURE IMAGING SYSTEM

Citing sources [18],[22], in a theoretical context, the upper bound on the attainable super-resolution magnification factor in a multi-aperture system is governed by several key factors. These factors encompass the diffraction limit associated with an individual aperture lens, the Nyquist angular frequency, and the aggregate count of apertures in the camera array. When limiting the analysis to these foundational variables and without incorporating factors such as registration precision, optical imperfections, and non-redundancy, the theoretical maximum super-resolution factor of a camera array imaging system can be computed using the subsequent formula.

$$f_{diffraction} = \frac{A}{1.22\lambda} \quad (1)$$

$$f_{Nyquist} = \frac{f}{2p} \quad (2)$$

$$R_{SR} = \min\left(\frac{f_{diffraction}}{f_{Nyquist}}, \sqrt{K}\right) = \min\left(\frac{2Ap}{1.22\lambda f}, \sqrt{K}\right) \quad (3)$$

In the Eq.(3), $f_{diffraction}$ represents the diffraction limit angular frequency, $f_{Nyquist}$ corresponds to the Nyquist angular frequency, $p$ indicates the pixel size of the camera, $A$ stands for the effective aperture diameter, and $\lambda$ refers to the wavelength. Within this setup, as depicted in Fig. 1, all nine apertures were employed using the Basler lens (C125-2522-5M-P, 25mm, F2.2-22) in conjunction with the monochromatic camera (Hikvision MV-CA004-10UM), featuring a pixel size of 6.9μm. The range of wavelengths considered for reconstruction in this study spans from 400nm to 700nm. When determining the theoretical upper limit of super-resolution magnification, a center wavelength of 700nm is the selected reference. Following Eq. (3), when each lens's aperture of the multi-aperture prototype imaging system is set to its maximum, $R_{SR}$ is determined as $\min(7.345, \sqrt{K})$, enabling the establishment of an upper limit for the theoretical super-resolution magnification factor at $\sqrt{K}$.

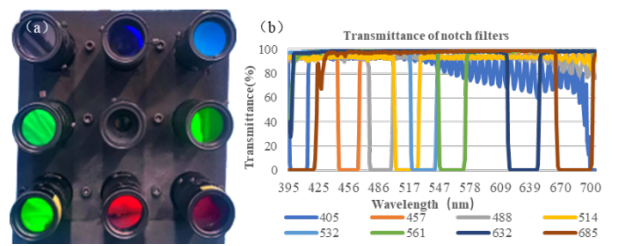

Fig. 1. (a) Proposed Multi-Aperture Multi-Spectral Imaging System and (b) Transmittance of the Selected Notch Filters.

## 2.1 The Imaging Model of the Notch Filters based Multi-aperture Imaging System

To enhance the resolution of images captured by the camera array in both spatial and spectral domains, we need to consider how to downsample spatial and spectral information when constructing the observation model for the camera array. The observation model can be formulated as follows:

$$Y_{L,i} = T_i X_H W_i R \quad (i=1,2...K) \quad (4)$$

$$Y_{H,i} = T_i X_H \quad (i=1,2...K) \quad (5)$$

$$Y_H = \begin{bmatrix} Y_{H,1} \\ Y_{H,2} \\ \vdots \\ Y_{H,9} \end{bmatrix} \quad (6)$$

$Y_{L,i} \in \mathbb{R}^{1 \times MN}$ represents the low-resolution notch-filtered or panchromatic image acquired by the $i_{th}$ aperture, $K$ represents the quantity of apertures within the camera array imaging system, $T_i \in \mathbb{R}^{1 \times Q}$ represents the spectral transmission curve of the $i_{th}$ selected notch filter. $Q$ represents the number of spectral images to be solved, $X_H \in \mathbb{R}^{Q \times KMN}$ represents the original high-resolution multi-spectral image datacube, $W_i \in \mathbb{R}^{KMN \times KMN}$ stands for the geometric transformation matrix, and $R \in \mathbb{R}^{KMN \times MN}$ represents the spatial downsampling factor. $Y_{H,i} \in \mathbb{R}^{1 \times KMN}$ depicts the high-resolution image, either notch-filtered or panchromatic, obtained through the $i_{th}$ aperture after registration with the central aperture.

In an ideal scenario, the spectral transmittance of the notch filters can be approximately expressed by the following formula:

$$T_i(\lambda) = \begin{cases} 1, & otherwise \\ 0, & |\lambda - a_i| \leq b_i \end{cases} \quad (7)$$

## 3 SPATIAL AND SPECTRAL SR ALGORITHM FOR NOTCH-FILTER BASED MULTI-APERTURE SYSTEM

The proposed multi-aperture prototype imaging system, equipped with 8 distinct notch filters, captures a total of 8 notch-filtered low-resolution images along with 1 panchromatic low-resolution image. These captured low-resolution images are then processed using the proposed SSR algorithm to reconstruct 31 high-resolution spectral images. The proposed SSR algorithm comprises three key steps: (1) Image registration, as detailed in reference [18], is conducted to determine $W_i$, which represents the geometric transformation matrix for each aperture in relation to the central aperture. Subsequently, the multi-aperture super-resolution algorithm is applied to reconstruct a high spatial resolution panchromatic image $\hat{Y}_{H,5}$, which consists of rich texture and intricate details. with rich spatial texture details. $\hat{Y}_{H,5}$ exhibits spectral disparities compared to $Y_{H,5}$ due to the non-uniform transmission of notch filters, while preserving consistent texture structure information. (2) Enhancement of spatial resolution for the notch-filtered images is accomplished by integrating structural details from $\hat{Y}_{H,5}$ through the application of a pan-sharpening algorithm. Although the use of notch filters affects the spectral characteristics of the reconstructed high-resolution $\hat{Y}_{H,5}$, the notch-filtered images contain relatively complete spatial details of the target scenes, making the spatial information of $\hat{Y}_{H,5}$ and $Y_{H,5}$ highly similar. Consequently, we are considering the utilization of pan-sharpening methods to generate the $Y_{H,i}$ images, which are expected to be highly accurate high-resolution notch-filtered images. (3) In the final step, the original high-resolution spectral images, referred to as $X_H$, are reconstructed using a spectral reconstruction algorithm, leveraging the high-resolution notch-filtered images and the panchromatic image. The entire algorithmic

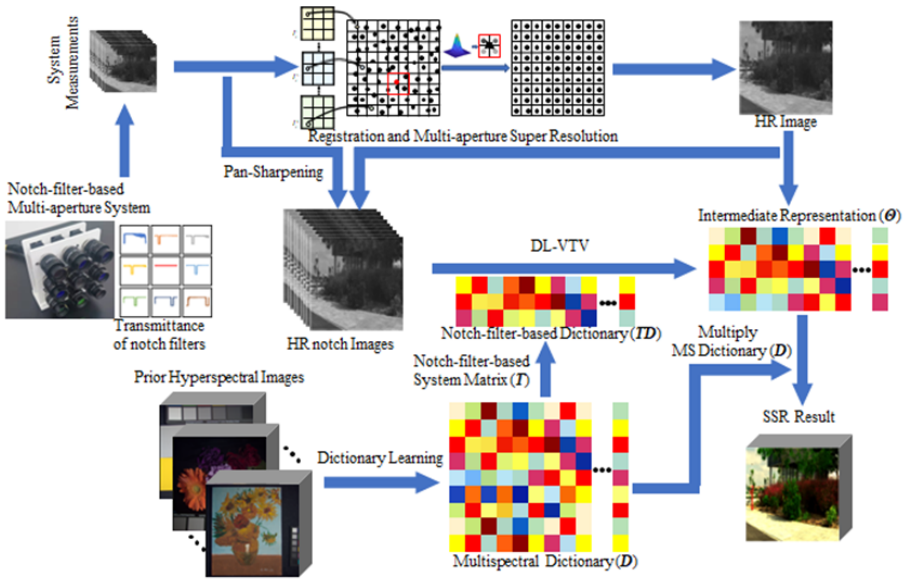

Fig. 2. Schematic of the Proposed Spatial and Spectral Super-Resolution Reconstruction Method

framework is depicted in Fig. 2.

### 3.1 3.1 Spatial Resolution Improvement

The system matrix $G \in \mathbb{R}^{KMN \times KMN}$ is derived from $W_i$, the geometric transformation matrix for the $i_{th}$ aperture relative to the central aperture. Despite its extensive dimension, $G$ contains numerous zero elements, forming a large sparse matrix that operates efficiently in terms of memory usage and computational speed

$$G = [W_1 R \; W_2 R \; \cdots \; W_9 R] \quad (8)$$

The classic image restoration model can be obtained from Eq.(8)

$$Y_L = \hat{Y}_{H,5} G, \quad (9)$$

where $Y_L$ is the composite of $Y_L^1 \sim Y_L^9$. Given the substantial dimension of the system matrix $G$, direct computation of its inverse becomes challenging, necessitating the adoption of an iterative approach for its solution. The corresponding cost function for this method is defined as follows:

$$E(\hat{Y}_{H,5}) = (\hat{Y}_{H,5} G - Y_L)^2 \quad (10)$$

Through the iterative approach of maximum likelihood estimation, a solution to Eq. (10) can be obtained incrementally. The iterative procedure is as follows:

$$\left(\hat{Y}_{H,5}^{n+1}\right)^T = diag\left(\left(\hat{Y}_{H,5}^{n}\right)^T\right) \\ * G\left\{diag\left(\left[G^T \left(\hat{Y}_{H,5}^{n}\right)^T\right]\right)\right\}^{-1} Y_L^T, \quad (11)$$

where $\hat{Y}_{H,5}^{n+1}$ represents the approximate result at the $n+1$ iteration, and $diag(\bullet)$ represents the diagonalization matrix of the vector. Eq. (11) can be employed to estimate the high-resolution image.

### 3.2 Improved Pansharpening Method fo Multi-Aperture Notch Filter Image Generating

The pan-sharpening approach fuses the high-resolution panchromatic image, $\hat{Y}_{H,5}$, rich in spatial detail, with the spectral resolution information contained in the low-resolution spectral images $Y_{L,1} \sim Y_{L,9}$. This fusion process yields high-resolution spectral images, designated as $Y_{H,1} \sim Y_{H,9}$, that effectively retain both spectral characteristics and spatial texture features. In this study, the pansharpening method described in [23] is predominantly utilized. The energy function of this method comprises two key terms, namely the Local Spectral Consistency term and the Dynamic Gradient Sparsity term. The first term, denoted as $E_1$, is focused on the preservation of spectral fidelity, as illustrated below:

$$E_1 = \frac{1}{2}\sum_{i=1}^{9}\|Y_{H,i} W_i R - Y_{L,i}\|_2 \quad (12)$$

Eq. (12) is deemed an ill-posed problem since it primarily focuses on ensuring the spectral faithfulness of the solution. Particularly, when a relatively large downsampling factor of 3 is employed, a significant loss of spatial details occurs, rendering it exceptionally challenging to find accurate solutions without imposing strong prior conditions. Fortunately, the panchromatic image serves as an excellent resource for obtaining precise high frequency information and boundary details. To successfully recover the missing spatial features in the solution, the inclusion of the following spatial fidelity components becomes essential:

$$E_2 = \|\nabla Y_H - \nabla P\|_{2,1}, \quad (13)$$

where $\nabla$ represents the gradient operator, and $P$ means duplicating $\hat{Y}_{H,5}$ to 9 bands. To exploit the inter-band correlation within each spectral band, the $l_{2,1}$ norm is introduced, aiming to encourage both dynamic gradient sparsity and group sparsity across the spectral domain. The specific solving procedure can be found in[23].

$$E = E_1 + E_2 = \frac{1}{2}\sum_{i=1}^{9}\|Y_{H,i} W_i R - Y_{L,i}\|_2 \\ + \lambda \|\nabla Y_H - \nabla P\|_{2,1}. \quad (14)$$

Due to the efficient convergence properties of the Fast Iterative Shrinkage-Thresholding Algorithm (FISTA) [24], the solution to Eq. (14) can be effectively obtained using its algorithmic framework. This equation can be decomposed into the following components:

$$I = I^j - \begin{bmatrix} (Y_{H,1} W_1 R - Y_{L,1})(W_1 R)^T \\ (Y_{H,2} W_2 R - Y_{L,2})(W_2 R)^T \\ \vdots \\ (Y_{H,9} W_9 R - Y_{L,9})(W_9 R)^T \end{bmatrix}, \quad (15)$$

$$Y_H^j = \arg\min_{Y_H}\left\{\frac{1}{2}\|Y_H - I\|_F^2 + \gamma\|\nabla Y_H - \nabla P\|_{2,1}\right\}, \quad (16)$$

where $j$ represents the $j_{th}$ iteration. Let $Z = Y_H - P$, which leads to the following equation:

$$Z^j = \arg\min_{Z}\left\{\frac{1}{2}\|Z - (I - P)\|_F^2 + \gamma\|\nabla Z\|_{2,1}\right\} \quad (17)$$

Eq. (17) is a vector total variation denoising model. $\|\nabla Z\|_{2,1} = \sum_{g=1}^{KMN}\sqrt{\sum_{i=1}^{9}((\nabla_1 Z_{i,g})^2 + (\nabla_2 Z_{i,g})^2)}$, where $\nabla_1$ and $\nabla_2$ represent the horizontal and vertical gradient operators, respectively. The solution methods for these operators are outlined in references [25],[26]. The update for $Y_H^j$ can be performed using the following formula:

$$Y_H^j = Z^j + P. \quad (18)$$

Once the solution for $Y_H^j$ is obtained, the update of $I^{j+1}$ within the FISTA framework is dependent on matrices $Y_H^j$ and $Y_H^{j-1}$. The step size $t$ for each iteration can be updated using the following formula:

$$t^{j+1} = (1 + \sqrt{1 + 4(t^j)^2})/2, \quad (19)$$

$$I^{j+1} = Y_H^j + \frac{t^j - 1}{t^{j+1}}(Y_H^j - Y_H^{j-1}), \quad (20)$$

where $t^1$ represents the initial step size for the iteration, in this particular problem $t^1 = 1$.

### 3.3 Dictionary learning and vector total variation spectral reconstruction

Through the combination of a dictionary learning-based spectral reconstruction algorithm with dynamic gradient sparsity and spectral sparsity constraints used in Pansharpening, the optimization objective is formulated as follows:

$$\arg\min_{\boldsymbol{\Theta}} \frac{1}{2}\|\boldsymbol{Y}_H - \boldsymbol{TD\Theta}\|_F^2 + \eta\|\boldsymbol{\Theta}\|_1 + \eta_{TV}\|\nabla \boldsymbol{X}_H - \nabla \mathrm{P}_Q\|_{2,1}, \quad (21)$$

$$\boldsymbol{T} = \begin{bmatrix} \boldsymbol{T}_1 \\ \boldsymbol{T}_2 \\ \vdots \\ \boldsymbol{T}_9 \end{bmatrix}, \quad (22)$$

where $\boldsymbol{D} \in \mathbb{R}^{Q \times 2Q}$ denotes a dictionary trained by K-single value decomposition algorithm[27]. $\boldsymbol{\Theta}$ denotes the sparse reprensation of $\boldsymbol{X}_H$, $\boldsymbol{X}_H = \boldsymbol{D\Theta}$. $\mathrm{P}_Q$ means duplicating $\hat{\boldsymbol{Y}}_{H,5}$ to $Q$ bands.

Eq. (21) is solved using the alternating direction method of multipliers (ADMM). By introducing auxiliary variables $\boldsymbol{Z}_1 = \boldsymbol{\Theta}$ and $\boldsymbol{Z}_2 = \boldsymbol{D\Theta}$ to Eq. (21), the corresponding augmented Lagrangian function is structured as follows:

$$\mathcal{L}(\boldsymbol{\Theta}, \boldsymbol{Z}_1, \boldsymbol{Z}_2, V_1, V_2) = \frac{1}{2}\|\boldsymbol{Y} - \boldsymbol{TD\Theta}\|_F^2 + \eta\|\boldsymbol{Z}_1\|_1$$
$$+ \eta_{TV}\|\nabla \boldsymbol{Z}_2 - \nabla \mathrm{P}_Q\|_{TV}$$
$$+ \frac{\rho_1}{2}\left\|\boldsymbol{\Theta} - \boldsymbol{Z}_1 + \frac{V_1}{\rho_1}\right\|_F^2 \quad (23)$$
$$+ \frac{\rho_2}{2}\left\|\boldsymbol{D\Theta} - \boldsymbol{Z}_2 + \frac{V_2}{\rho_2}\right\|_F^2,$$

where $V_1$ and $V_2$ are the Lagrange multipliers, and $\rho_1$ and $\rho_2$ are the coefficients of the regular terms. Following to ADMM, iteratively solve variables $\boldsymbol{\Theta}$, $\boldsymbol{Z}_1$, $\boldsymbol{Z}_2$, $V_1$ and $V_2$ until convergence is achived.

$$\boldsymbol{\Theta}^{k+1} = \arg\min_{\boldsymbol{\Theta}} \mathcal{L}(\boldsymbol{\Theta}, \boldsymbol{Z}_1^k, \boldsymbol{Z}_2^k, V_1^k, V_2^k), \quad (24)$$

$$\boldsymbol{\Theta}^{k+1} = (\boldsymbol{D}^T \boldsymbol{T}^T \boldsymbol{TD} + \rho_1 \boldsymbol{I} + \rho_2 \boldsymbol{D}^T \boldsymbol{D})^{-1} \times$$
$$\left[\boldsymbol{D}^T \boldsymbol{T}^T \boldsymbol{Y} + \rho_1\left(\boldsymbol{Z}_1^k - \frac{V_1^k}{\rho_1}\right) + \rho_2\left(\boldsymbol{Z}_2^k - \frac{V_2^k}{\rho_2}\right)\right], \quad (25)$$

$$\boldsymbol{Z}_1^{k+1} = \arg\min_{\boldsymbol{Z}_1} \mathcal{L}(\boldsymbol{\Theta}_1^{k+1}, \boldsymbol{Z}_1, V_1^k), \quad (26)$$

$$\boldsymbol{Z}_1^{k+1} = soft\left(\boldsymbol{\Theta}^{k+1} + \frac{V_1^k}{\rho_1}, \frac{\eta}{\rho_1}\right), \quad (27)$$

$$\boldsymbol{Z}_2^{k+1} = \arg\min_{\boldsymbol{Z}_2} \mathcal{L}(\boldsymbol{\Theta}_1^{k+1}, \boldsymbol{Z}_2, V_2^k), \quad (28)$$

$$\boldsymbol{Z}_2^{k+1} = \arg\min_{\boldsymbol{Z}_2} \frac{\rho_2}{2}\left\|\boldsymbol{D\Theta}^{k+1} - \boldsymbol{Z}_2 + \frac{V_2^k}{\rho_2}\right\|_F^2 \quad (29)$$
$$+ \eta_{TV}\|\nabla \boldsymbol{Z}_2 - \nabla \mathrm{P}_Q\|_{1,2},$$

where $k$ represents $k_{th}$ iteration. Let $\boldsymbol{Z}_Q = \boldsymbol{Z}_2 - \mathrm{P}_Q$, and Eq. (29) can be rewritten as:

$$\boldsymbol{Z}_Q^{k+1} = \arg\min_{\boldsymbol{Z}_2} \frac{\rho_2}{2}\left\|\boldsymbol{D\Theta}^{k+1} - (\boldsymbol{Z}_Q - \mathrm{P}_Q) + \frac{V_2^k}{\rho_2}\right\|_F^2 \quad (30)$$
$$+ \eta_{TV}\|\nabla \boldsymbol{Z}_Q\|_{1,2},$$

$$\boldsymbol{Z}_2^{k+1} = \boldsymbol{Z}_Q^{k+1} + \mathrm{P}_Q. \quad (31)$$

Eq. (31) is a vector total variation model. The Lagrange multipliers are updated by:

$$V_1^{k+1} = V_1^k + \rho_1(\boldsymbol{\Theta}^{k+1} - \boldsymbol{Z}_1^{k+1}), \quad (32)$$
$$V_2^{k+1} = V_2^k + \rho_2(\boldsymbol{D\Theta}^{k+1} - \boldsymbol{Z}_2^{k+1}). \quad (33)$$

## 4 SIMULTION

The effectiveness of the proposed algorithm was verified using the CAVE dataset [28], which consists of 31 indoor spectral images. Each image is composed of 31 spectral bands and has a resolution of 512×512 pixels, covering a wavelength range from 400nm to 700nm with a spectral resolution of 10nm. To align with the observation model of the Notch Filters based Multi-aperture Imaging System, the original spectral images extracted from the CAVE dataset were subjected to geometric transformations and downsampling, resulting in the generation of $Y_{L,1} \sim Y_{L,9}$.

To enable a meaningful comparative analysis, the data scales between our proposed system and CASSI were aligned. The initial high-resolution spectral image, cropped from the CAVE dataset, had dimensions of 510×510×31. Subsequently, a downsampling factor of 3 was applied, resulting in the generation of simulated notch-filtered spectral images measuring 170×170×9. This led to a compression ratio of 31:1 for the proposed system. In contrast, CASSI's captured image featured dimensions of 512×542 pixels, and the reconstructed image size was 512×512×31, yielding a compression ratio of approximately 29.3:1. This consistency in data scales ensured a convenient and meaningful comparative analysis. CASSI is widely recognized as a reliable computational spectral imaging system, and its algorithm has undergone extensive improvements over the years, reaching a high level of performance. Hence, to assess the effectiveness of the SSR algorithm, a comparative evaluation was conducted with CASSI[13] through simulation. The simulation employed the coded aperture mask as reported in [29], resulting in a coded image with dimensions of 512×542 pixels, a size similar to the total pixel count of our proposed system. The reconstruction algorithms used for CASSI included TwIST+TV[30], GAP-TV[31], and PnP[29]. In TwIST, the iteration number was set to 150, and the regularization parameter was fixed at τ = 0.1. The iterations and parameters for PnP and GAP-TV were configured as per the specifications in [29]. The parameters for the proposed SSR method were configured as follows: multi-aperture super-resolution iteration $n = 50$, pansharpening parameters $L = 1$, $t^1 = 1$, $\gamma = 0.05$, $Jter_{max} = 200$, spectral reconstruction parameters $\rho_1 = 0.000001$, $\rho_2 = 0.002$, $\eta_{TV} = 0.001$, $\eta = 0.001$, $Iter_{max} = 40$. All algorithms were executed on a laptop with an AMD R5-4600H (3.0GHz) processor and 16GB of memory. TwIST+TV and SSR were implemented in MATLAB, while GAP-TV and PnP were developed in Python.

The results of our simulations conducted with the CAVE dataset are presented in Tables 1 and 2. In Table 1, details on the Peak Signal-to-Noise Ratio (PSNR) and Structural Similarity Index (SSIM) for the reconstruction of 15 Multi-spectral Images (MSIs) from the CAVE dataset are provided. The data in Table 1 clearly indicates that CASSI yields lower PSNR and SSIM values compared to our proposed system, which achieves a significant advantage. The PnP method is outperformed by our system by a margin of 5dB in PSNR, and SSIM is increased by 0.1. This phenomenon can be attributed to the coded mask used in the CASSI system, as its imaging mechanism leads to significant degradation of the spatial structure, making it challenging to preserve the original spatial information in the

TABLE 1

PSNR AND SSIM VALUES FOR THE 15 RECONSTRUCTED MSIS FROM THE CAVE DATASET. THE AVERAGE VALUES ARE CALCULATED AND LISTED AT THE BOTTOM

| HSI | RMSE | | | | SAM | | | |
|---|---|---|---|---|---|---|---|---|
| | TwIST | GAP-TV | PnP | SSR | TwIST | GAP-TV | PnP | SSR |
| Balloons | 6.488 | 5.407 | 4.991 | **3.536** | 0.090 | **0.085** | 0.119 | 0.128 |
| Beads | 23.553 | 20.733 | 20.740 | **9.534** | 0.304 | 0.310 | 0.310 | **0.207** |
| CD | 8.528 | **6.463** | 6.829 | 7.727 | 0.121 | 0.134 | 0.193 | 0.197 |
| Toy | 16.290 | 14.806 | 14.446 | **6.587** | 0.182 | 0.187 | 0.210 | **0.119** |
| Clay | 6.684 | 5.899 | 5.872 | **3.632** | 0.190 | 0.228 | 0.315 | **0.178** |
| Cloth | 17.657 | 17.145 | 16.431 | **9.491** | 0.170 | 0.171 | 0.171 | **0.137** |
| Egyptian | 5.796 | 5.789 | 6.049 | **2.715** | 0.264 | 0.256 | 0.347 | **0.218** |
| Face | 6.050 | 6.092 | 6.059 | **3.563** | **0.129** | 0.144 | 0.244 | 0.146 |
| Beers | 7.888 | 6.340 | 4.956 | **4.341** | **0.044** | 0.041 | 0.042 | 0.061 |
| Food | 7.895 | 7.086 | 6.334 | **5.300** | 0.153 | 0.181 | 0.213 | **0.152** |
| Lemon | 9.294 | 8.376 | 7.488 | **3.558** | 0.176 | 0.229 | 0.253 | **0.147** |
| Lemons | 5.435 | 5.559 | 5.624 | **2.835** | **0.106** | 0.117 | 0.217 | 0.145 |
| Peppers | 9.646 | 7.947 | 7.548 | **3.703** | 0.156 | 0.165 | 0.241 | **0.143** |
| Strawberries | 6.392 | 6.947 | 7.415 | **3.858** | 0.148 | 0.177 | 0.248 | **0.131** |
| Sushi | 6.701 | 6.231 | 5.813 | **3.334** | **0.105** | 0.126 | 0.184 | 0.163 |
| Average | 9.620 | 8.721 | 8.440 | **4.914** | 0.156 | 0.170 | 0.220 | **0.151** |

TABLE 2

RMSE AND SAM VALUES FOR THE 15 RECONSTRUCTED MSIS FROM THE CAVE DATASET. THE AVERAGE VALUES ARE CALCULATED AND LISTED AT THE BOTTOM

| HSI | RMSE | | | | SAM | | | |
|---|---|---|---|---|---|---|---|---|
| | TwIST | GAP-TV | PnP | SSR | TwIST | GAP-TV | PnP | SSR |
| Balloons | 6.488 | 5.407 | 4.991 | **3.536** | 0.090 | **0.085** | 0.119 | 0.128 |
| Beads | 23.553 | 20.733 | 20.740 | **9.534** | 0.304 | 0.310 | 0.310 | **0.207** |
| CD | 8.528 | **6.463** | 6.829 | 7.727 | 0.121 | 0.134 | 0.193 | 0.197 |
| Toy | 16.290 | 14.806 | 14.446 | **6.587** | 0.182 | 0.187 | 0.210 | **0.119** |
| Clay | 6.684 | 5.899 | 5.872 | **3.632** | 0.190 | 0.228 | 0.315 | **0.178** |
| Cloth | 17.657 | 17.145 | 16.431 | **9.491** | 0.170 | 0.171 | 0.171 | **0.137** |
| Egyptian | 5.796 | 5.789 | 6.049 | **2.715** | 0.264 | 0.256 | 0.347 | **0.218** |
| Face | 6.050 | 6.092 | 6.059 | **3.563** | **0.129** | 0.144 | 0.244 | 0.146 |
| Beers | 7.888 | 6.340 | 4.956 | **4.341** | **0.044** | 0.041 | 0.042 | 0.061 |
| Food | 7.895 | 7.086 | 6.334 | **5.300** | 0.153 | 0.181 | 0.213 | **0.152** |
| Lemon | 9.294 | 8.376 | 7.488 | **3.558** | 0.176 | 0.229 | 0.253 | **0.147** |
| Lemons | 5.435 | 5.559 | 5.624 | **2.835** | **0.106** | 0.117 | 0.217 | 0.145 |
| Peppers | 9.646 | 7.947 | 7.548 | **3.703** | 0.156 | 0.165 | 0.241 | **0.143** |
| Strawberries | 6.392 | 6.947 | 7.415 | **3.858** | 0.148 | 0.177 | 0.248 | **0.131** |
| Sushi | 6.701 | 6.231 | 5.813 | **3.334** | **0.105** | 0.126 | 0.184 | 0.163 |
| Average | 9.620 | 8.721 | 8.440 | **4.914** | 0.156 | 0.170 | 0.220 | **0.151** |

reconstructed spectral images. In Table 2, the results for Root Mean Square Error (RMSE) and Spectral Angle Mapper (SAM) are presented for the reconstruction of the same 15 MSIs from the CAVE dataset. The data reveals that the proposed system demonstrates a substantial reduction in RMSE and marginally outperforms the suboptimal algorithm in terms of SAM.

Fig. 3 features a comparison of spectral curve reconstructions accomplished through various methods. It highlights the contrast between reference spectral curves and reconstructed spectral curves across nine distinct areas (a-i) within three different spectral images ('food,' 'toy,' and 'cloth') sourced from the CAVE dataset. The results clearly indicate that, in numerous regions, the spectral curves reconstructed by the proposed system and algorithm outperform those generated by three different CASSI algorithms, namely TwIST[30], GAP-TV[31], and PnP[29]. Fig. 4 presents an in-depth analysis of the reconstruction results for the 'beads' spectral image from the CAVE dataset, with a specific focus on central wavelengths of 440nm, 540nm, and 600nm. Rows 1, 3, and 5 reveal the reconstructed images and corresponding reference images at different wavelengths, while rows 2, 4, and 6 display the associated error maps. Figure 5 provides insights into the reconstruction outcomes for 'cloth' spectral images from the CAVE dataset at central wavelengths of 450nm, 520nm, and 660nm. In addition, Figure 6 unveils the results for 'toy' spectral images at central wavelengths of 440nm, 540nm, and 600nm.

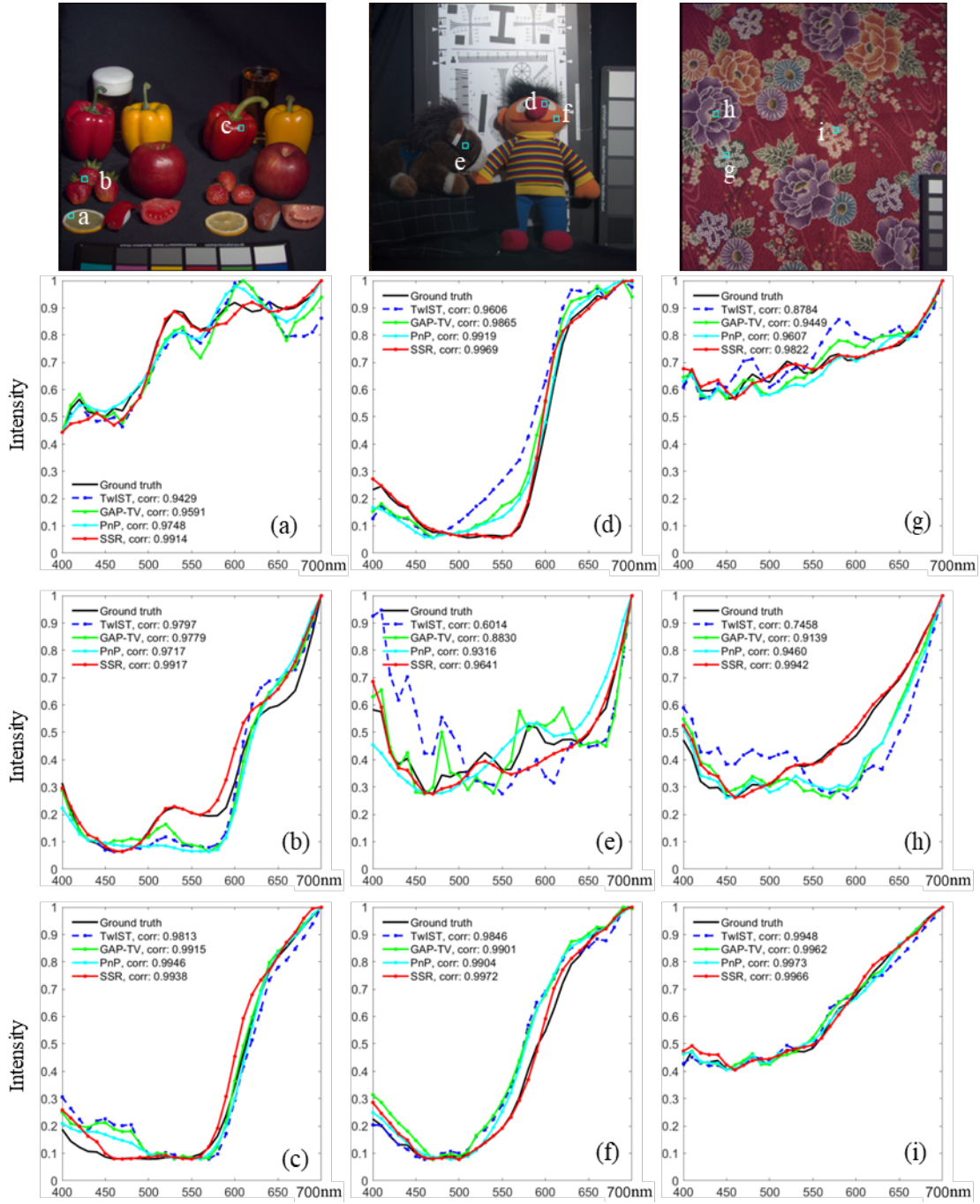

Fig. 3. Reconstructed spectral curves within nine designated regions (a)-(i) across three distinct spectral images ('food,' 'toy,' and 'cloth') from the CAVE dataset

As a collective observation, Figures 4-6 clearly demonstrate that the overall reconstruction error achieved by the proposed method is notably lower when compared to the CASSI methods. A prime example is found in Figure 6, where the CASSI system and PnP method can adequately reconstruct the toy's face while still grappling with the challenge of reproducing texture details across each spectral band for the background target. In contrast, the proposed SSR algorithm significantly enhances the spatial detail information across the entire image faithfully.

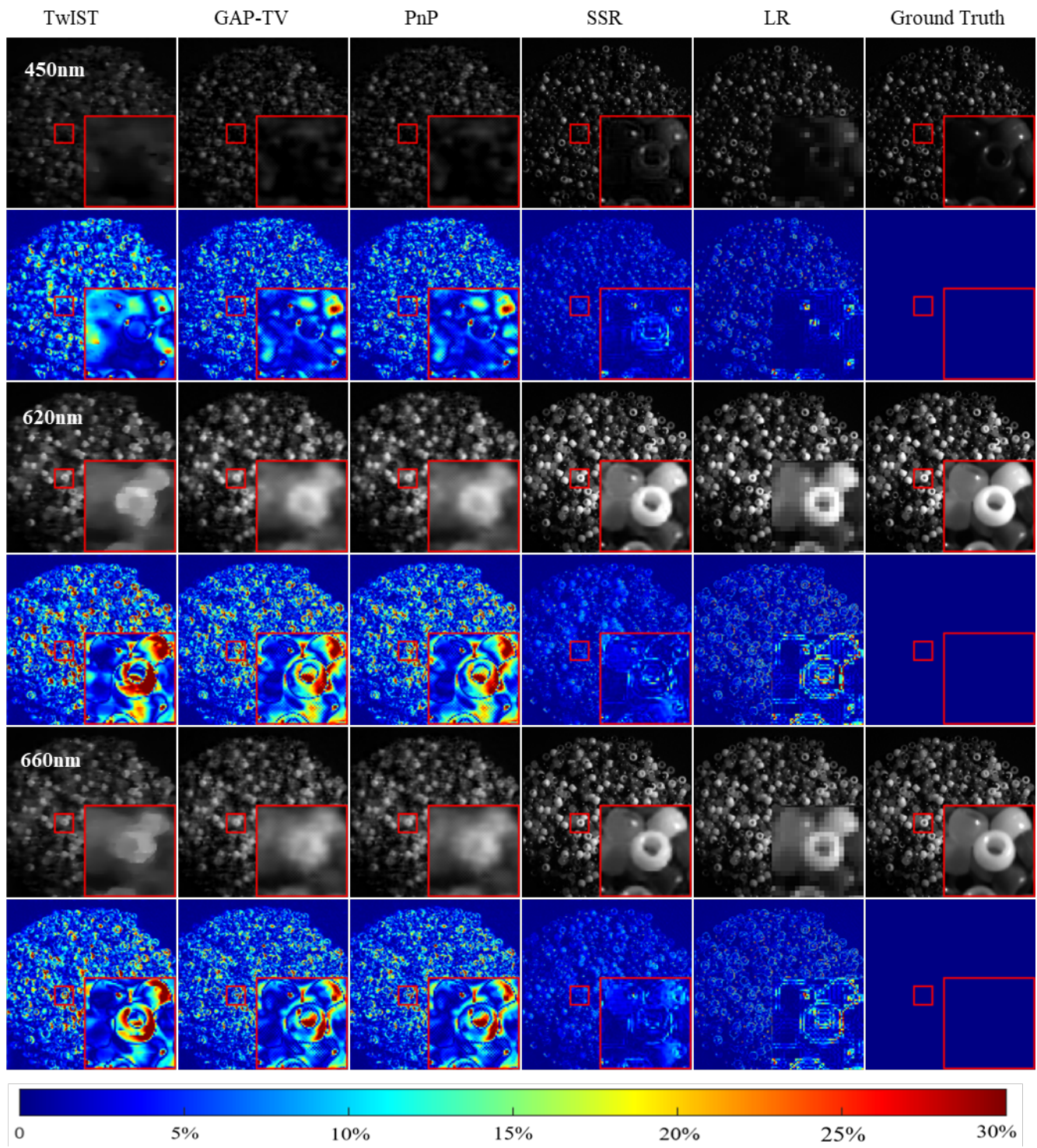

Fig. 4. Reconstruction results of the 'beads' spectral image from the CAVE dataset at central wavelengths of 450nm, 620nm, and 660nm. Rows 1, 3, and 5 display the reconstruction results, reference images, and magnified sections, while rows 2, 4, and 6 show the corresponding error maps

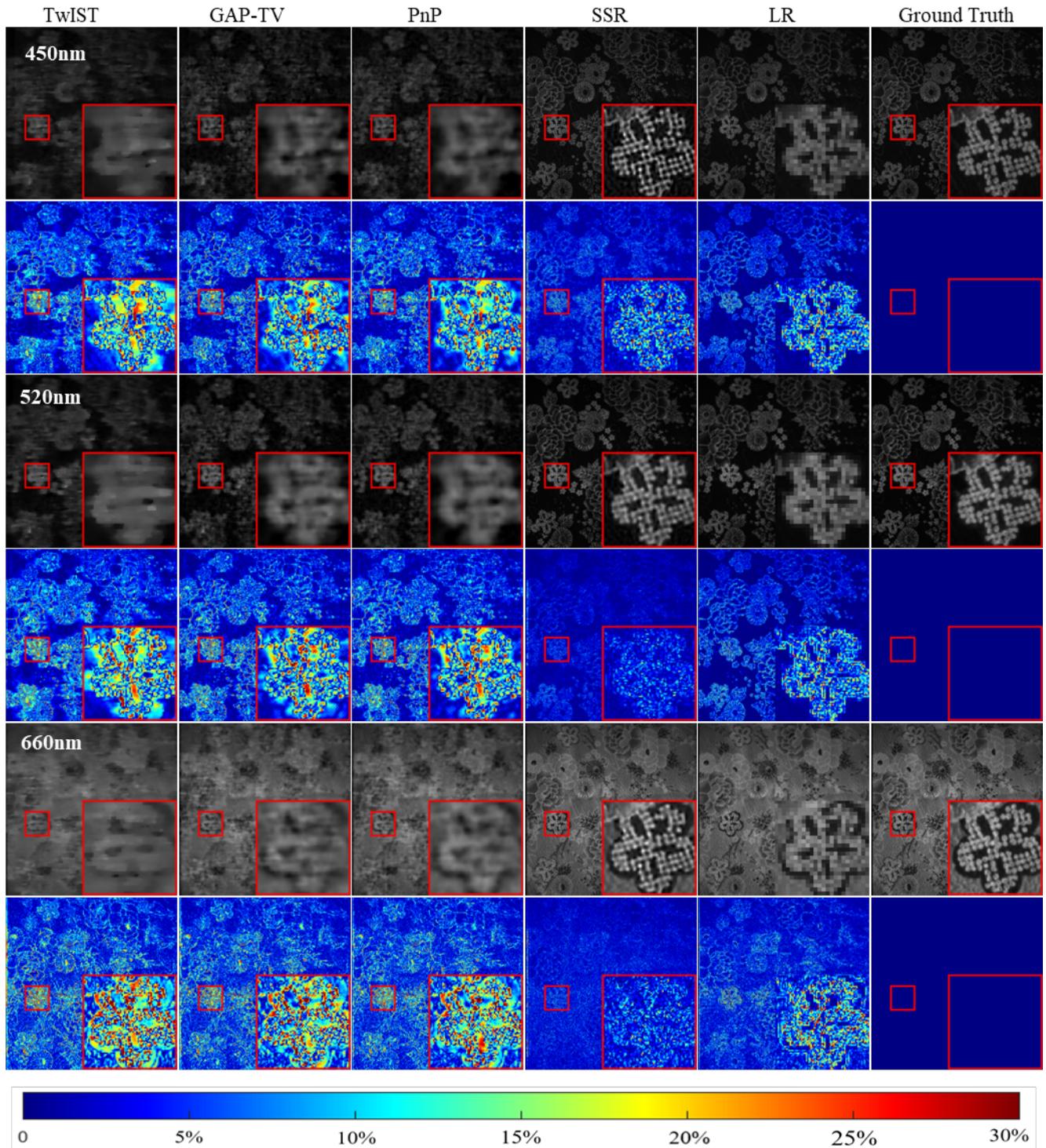

Fig. 5. Reconstruction results of the 'cloth' spectral image from the CAVE dataset at central wavelengths of 450nm, 520nm, and 660nm. Rows 1, 3, and 5 display the reconstruction results, reference images, and magnified sections, while rows 2, 4, and 6 show the corresponding error maps

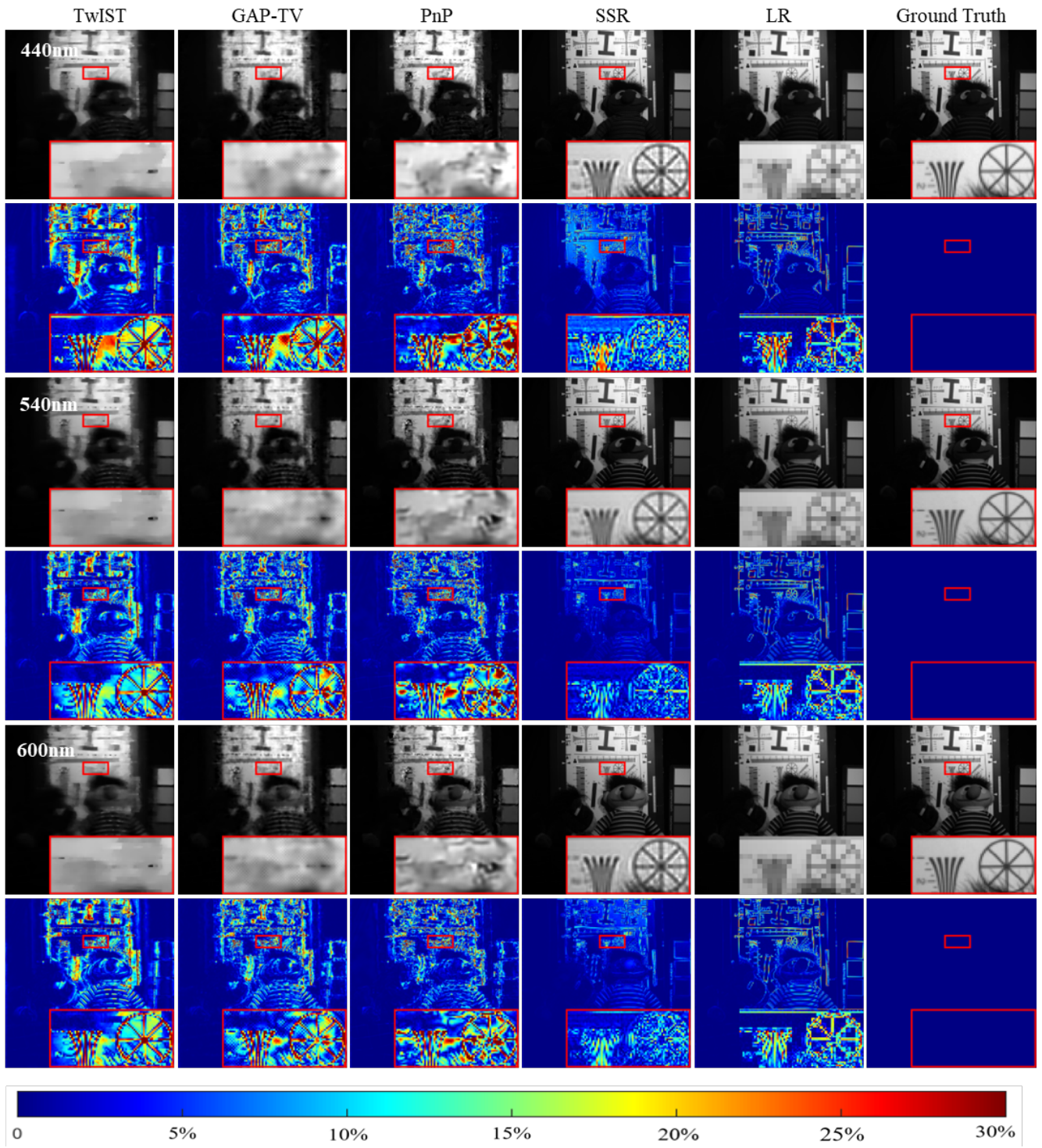

Fig. 6. Reconstruction results of the 'toy' spectral image from the CAVE dataset at central wavelengths of 440nm, 540nm, and 600nm. Rows 1, 3, and 5 display the reconstructed spectral images, reference images, and magnified sections, while rows 2, 4, and 6 show the corresponding error maps

TABLE 3
COMPARISON OF RECONSTRUCTION TIME ACROSS DIFFERENT METHODS

| Method | TwIST | GPA-TV | PnP | SSR |
|---|---|---|---|---|
| Time (s) | 500.5 | 569.8 | 540.8 | 173.4 |

## 4.1 Experiments using Actual Captured Data

The proposed prototype system comprises nine panchromatic cameras, each with a resolution of 540×720. This prototype is employed to capture images of an X-rite ColorChecker, a color box, and an ISO 12233 2014 eSFR target. In a single capture session, we obtain eight low-resolution notch images and one low-resolution panchromatic image. These images are then processed using the SSR algorithm introduced in this study to reconstruct a super-resolution spectral image with a spatial resolution of 1620×2160. The illumination source remains a spectral tunable light source (Thouslite LEDCube) with a spectral range spanning from 420nm to 690nm. Consequently, the spectral range of the captured images in this experiment also covers 420nm to 690nm, totaling 28 spectral bands. To simulate outdoor light conditions, the maximum illumination intensity of the light source is utilized. The camera's exposure time for capturing notch and panchromatic images is set at 0.4ms, resulting in a theoretical frame rate of 2500fps. Fig. 7 pro- obtained from the hyperspectral camera. Fig. 7 also presents the correlation coefficient between the reconstructed and reference spectral curves, along with the reconstruc-vides a comparison between the reconstructed spectral curves of 18 color blocks and the reference spectral curve tion results for all bands spanning from 420nm to 690nm.
To compare the SR MSIs generated using the proposed system and algorithms with the state-of-art single-frame super-resolution, the low-resolution panchromatic image of the center aperture was processed using single-frame super-resolution, employing a 3x super-resolution factor achieved through the convolutional neural network-based IMDN (a lightweight information multi-distillation network) [32]. In Fig. 8, you will find the RGB image of the 'color box' scene, the low-resolution panchromatic image captured by the center aperture, and a total of 28 high-resolution spectral images covering the 420-690nm range, generated through the SSR algorithm. Fig. 9 presents the reconstruction results of the 'ISO 12233' target. Due to its predominantly black and white areas, where each spectral band exhibits similar spectral reflectance, the focus is on illustrating the 540nm reconstruction results. The image produced by the IMDN algorithm generally appears brighter and handles text more effectively. However, it introduces some noise and results in erroneous super-resolution imaging outcomes for certain parts of the image, such as incorrect text and an inaccurate representation of the animated dog on the color box. In contrast, the SSR algorithm excels in recovering spatial details of the target scene without introducing image artifacts.

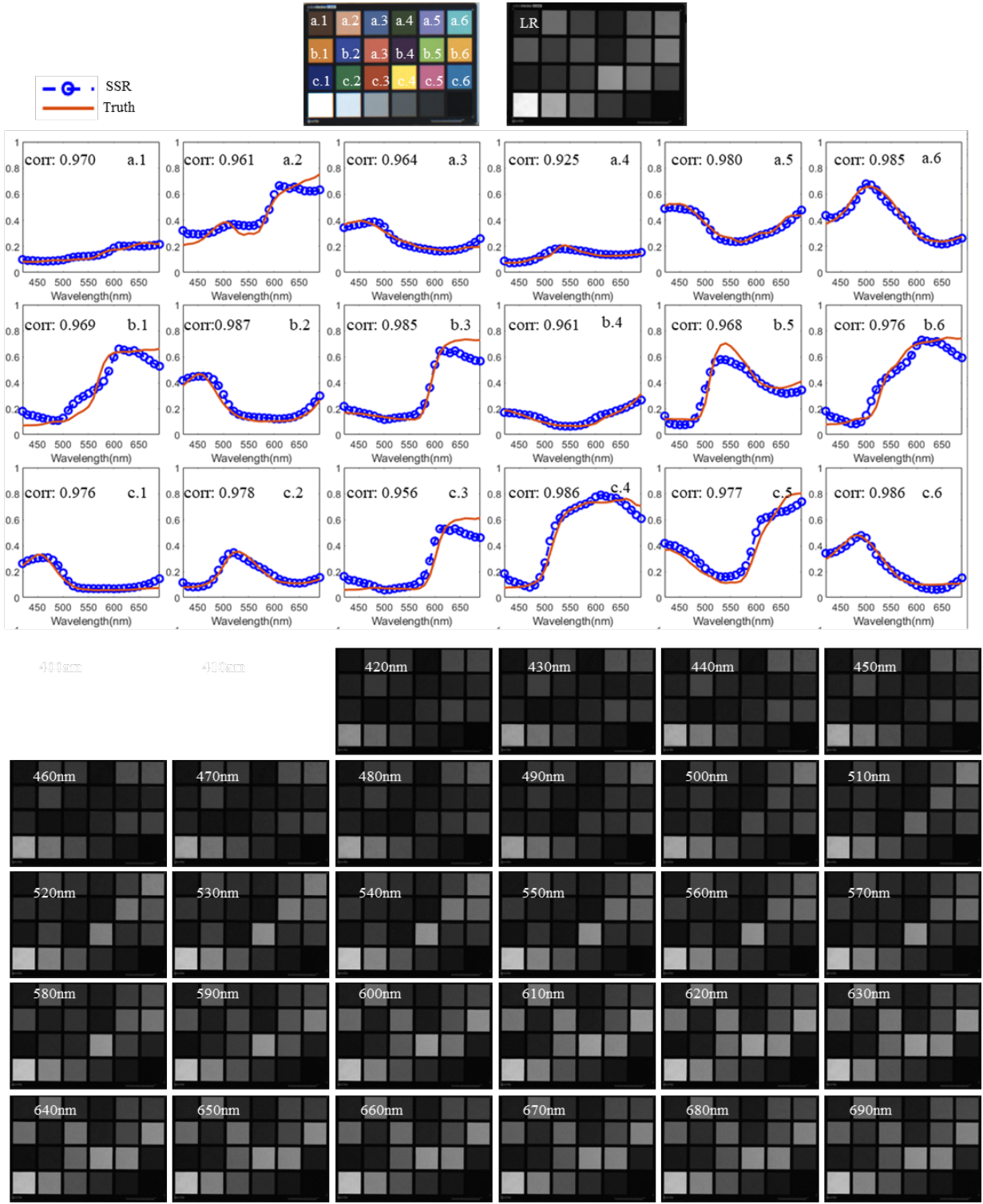

Fig. 7. Comparison of Reconstructed Spectral Curves and Ground Truth for the 18 Color Blocks. The reconstructed Multispectral Images (MSIs) are displayed at the bottom

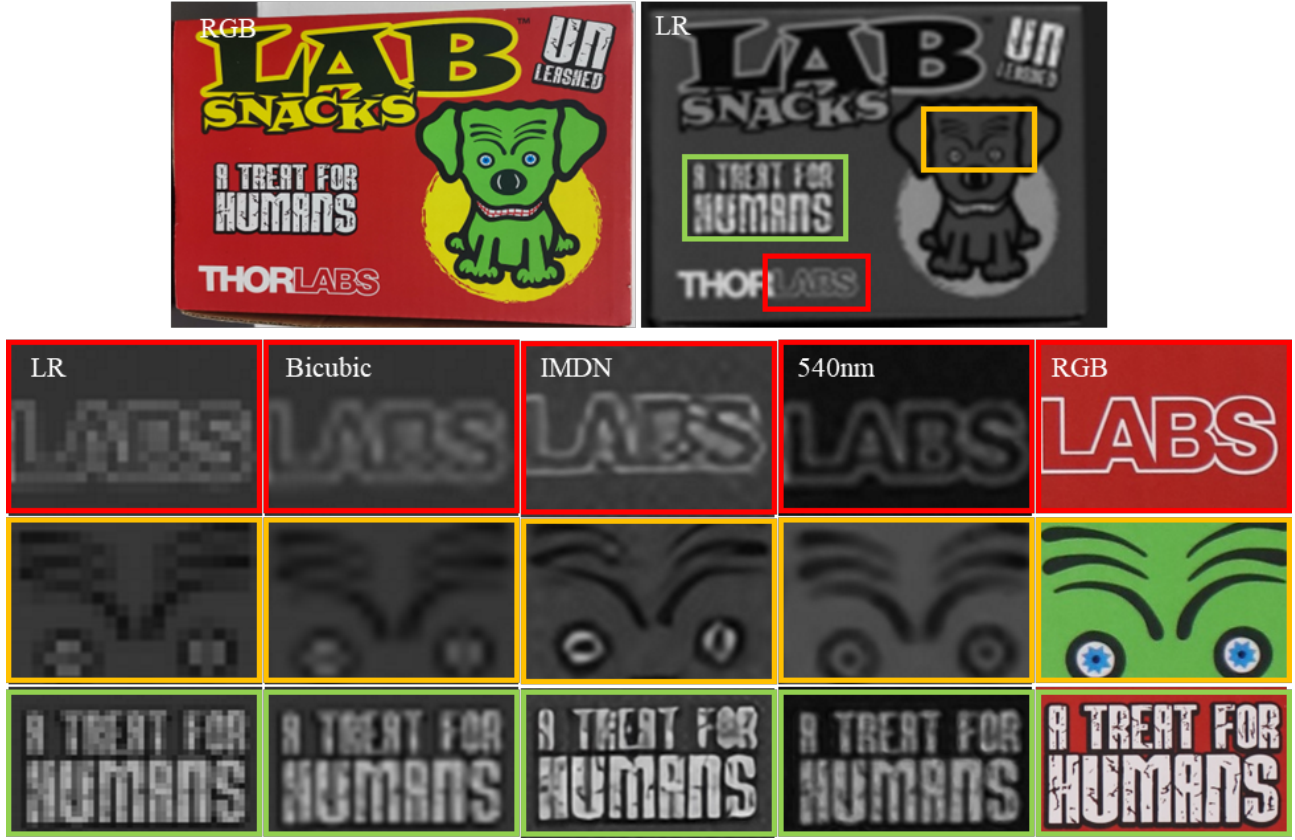

Fig. 8. A Comparison of Reconstructed High-Resolution Spectral Images for the Color Box. Rows 1-6 display 28 high-resolution spectral images reconstructed using SSR. Rows 7-9 show magnified images of three selected areas, presented sequentially from left to right: low-resolution panchromatic image, panchromatic image with bicubic interpolation, IMDN, SSR reconstructed image at 540nm, and the RGB image of the color box

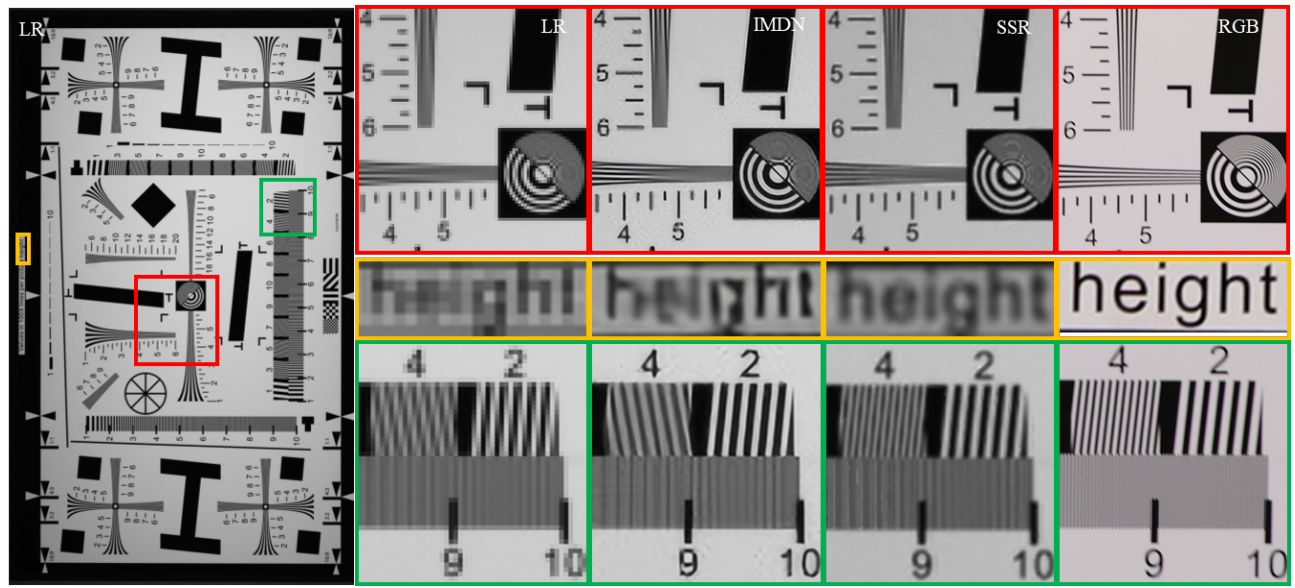

Fig. 9. A Comparison of Reconstructed High-Resolution Spectral Images for the Color Box. Rows 1-6 display 28 high-resolution spectral images reconstructed using SSR. Rows 7-9 show magnified images of three selected areas, presented sequentially from left to right: low-resolution panchromatic image, panchromatic image with bicubic interpolation, IMDN, SSR reconstructed image at 540nm, and the RGB image of the color box

## 5 CONCLUSION

In conclusion, this paper introduced a novel spatial and spectral super-resolution reconstruction method based on a notch filter multi-aperture imaging system, representing a significant advancement in enhancing both spatial and spectral resolution in multi-aperture system-acquired images. Leveraging the high light efficiency of the notch filters, the proposed system achieves an impressive imaging frame rate, showing promise for high-resolution spectral imaging of dynamic scenes with exceptional frame rates. The proposed algorithm effectively leverages complementary information across apertures for super-resolution reconstruction and subsequently recovers high-resolution spectral details of the target scene through a combination of pan-sharpening and spectral reconstruction methods. Simulation evaluations conducted using the CAVE dataset demonstrate that the proposed system and algorithm outperform the CASSI system in terms of both reconstruction quality and speed. Experimental verification further underscores the system's capabilities in spatial super-resolution and spectral fidelity. Future work will optimize the algorithm, employ an end-to-end neural network to enhance imaging quality, and enable real-time multispectral reconstruction at video frame rates.